\documentclass[aps,preprint]{revtex4}
\usepackage{ams,amsmath,color}
\usepackage[ansinew]{inputenc} % for accents
\usepackage{verbatim}
\usepackage{graphicx}

\begin{document}

\title{Carrier relaxation in Si/SiO$_2$ quantum dots}

\author{A.~A.~Prokofiev$^1$,
S.~V.~Goupalov$^{1,2}$,
A.~S. Moskalenko$^3$\footnote{Also at: \textit{Ioffe Physico-Technical Institute, Russian Academy of Sciences,
Polytechnicheskaya 26, 194021 St. Petersburg, Russia.}},
A.~N.~Poddubny$^1$,
I.~N.~Yassievich$^1$}
\affiliation{$^1$Ioffe Physico-Technical Institute, Russian Academy of Sciences \\
Polytechnicheskaya 26, 194021 Saint-Petersburg, Russia
\\
$^2$Department of Physics, Jackson State University, Jackson, MS 39217, USA
\\
$^3$Institut f\"ur Physik, Martin-Luther-Universit\"at Halle-Wittenberg,
Nanotechnikum-Weinberg, Heinrich-Damerow-St. 4, 06120 Halle, Germany}

\begin{abstract}
Carrier relaxation due to both optical and nonradiative intraband transitions
in silicon quantum dots (QDs) in SiO$_2$ has been considered.
Interaction of confined holes with optical phonons has been studied.
The Huang-Rhys factor is calculated for such transitions.
The probability of intraband transition of a confined hole emitting several optical phonons is estimated.
\end{abstract}

\maketitle

\section{Introduction}
\label{sec:intro}

The system of silicon quantum dots in silicon oxide matrix has been of great interest during the past decade.
Apart from increasing excitation rate of implanted erbium ions~\cite{Kenyon_PQE_2002}
Si nanocrystals can be used as a light source~\cite{Pavesi_Nature_2000}
or a photon cutting medium in photovoltaics applications~\cite{Timmerman_Nature_2008}.
The theory of single-particle states in spherical Si/SiO$_2$ quantum dots
has been recently developed~\cite{Moskalenko_PRB_2007}.
It was shown that most of the gaps between the calculated levels appeared to be large enough
to suppress single-phonon relaxation processes in QD with diameter in the range $2 \div 4$~nm.
Here we present the calculation of relaxation probabilities for confined electrons and holes.
Both --- optical and nonradiative --- intraband transitions are considered.
Multiphonon intraband transitions of valence-band holes
are due to their interaction with optical phonons via the deformation potential.
This mechanism of carrier-phonon coupling is only efficient for valence-band holes
while for conduction-band electrons it is forbidden by the crystal symmetry~\cite{BirPikus_book}.

Values of the Huang-Rhys parameter are calculated, and the probabilities of the transitions are estimated.

\section{Carriers confined in spherical Si quantum dots}
\label{sec:states}

Here we use the effective mass approach to spherical silicon nanocrystals developed in~\cite{Moskalenko_PRB_2007}.
The spin-orbit interaction is neglected in the consideration.
The details of the approximation are given below.

Electron envelope functions are found as a result of a numerical solution to the Schr\"{o}dinger equation
after separating the angular part $\exp(im\phi)$ ($m$ can be any integer number) as there is a strong anisotropy
of the electron effective mass.
Due to the conduction band symmetry the electron states are sixfold degenerate for m=0
and 12 times degenerate for $|m|>0$, as two opposite values $m=\pm|m|$ correspond to the same energy.
So the states are marked with $Ne_{|m|}$ where the letter $e$ shows that it is an
electron state, $N$ is the main quantum number which shows the order of the energy level for given
$|m|$.  For example, the ground state is marked as $1e_0$, which means that this is the first
electron state with $m=0$.

The hole states are characterized by the total angular momentum $F$ and parity.
The states with the parity $(-1)^F$ are the heavy hole ($hh$) states
while the states with the parity $(-1)^{F+1}$ are contributed by both the light and the heavy hole states
(mixed hole states: $hm$)
with the only exception for $F=0$ when the hole states are only contributed by the light holes ($hl$).
The hole states are described by the main quantum number $n=1,2,3...$
followed by the abbreviation $hh$, $hm$ or $hl$
and the subscript denoting the total angular momentum~$F$.
For example, the hole state with the lowest energy is of the mixed type --- $1hm_1$.

\begin{figure}
    \includegraphics[width=12cm]{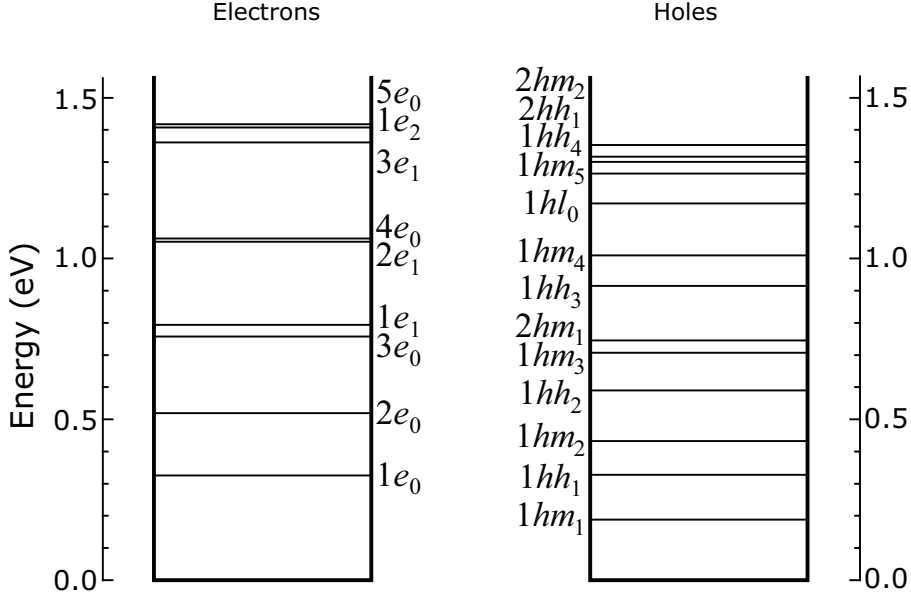}
\caption{
    \label{fig:levels31}
    Electron and hole energy levels in 3.1~nm quantum dot
}
\end{figure}

Calculated electron and hole energy levels are shown in Fig.~\ref{fig:levels31}.

\section{Optical intraband transitions of confined carriers}
\label{sec:optical}

Previously, the oscillator strengths were calculated for electrons in Si QD~\cite{Allan_PRB_2002}.
However, the tight-binding model used there greatly overestimates the carrier quantization energies
if applied to Si QD embedded into SiO$_2$.
Here we use our theory developed for carriers confined in Si/SiO$_2$ QD,
giving much better agreement with the experiments on this system,
and produce calculations for intraband transitions of holes as well.

The spontaneous emission rate is given by the relation between Einstein coefficients~\cite{Goupalov_PRB_2003}:
\begin{equation}
    \frac{1}{\tau_{\mathrm{rad}}}
    =
    \frac{e^2}{3 \pi \epsilon_0}
    \frac{n_{\mathrm{eff}}}{\hbar^4 c^3}(\hbar \omega_{ij})^3
    \frac{1}{N_i}
    \sum_{\mu_i \mu_j}
        \left|
            \langle
            j, \mu_j
            |
            \vec{r}
            |
            i, \mu_i
            \rangle
        \right|^2.
\label{A_via_f}
\end{equation}
Here $n_{\mathrm{eff}}$ is the effective refractive index of the media
given by the formulae below~\cite{Thraendhardt_PRB_2002}:
\begin{equation}
    n_{\mathrm{eff}}
    =
    \left(
        \frac{\kappa_{\mathrm{SiO_2}}}{\kappa_{\mathrm{eff}}}
    \right)^2
    \kappa_{\mathrm{SiO_2}}^{1/2},
\end{equation}
\begin{equation}
    \kappa_{\mathrm{eff}}
    =
    \frac{\kappa_{\mathrm{Si}}+2\kappa_{\mathrm{SiO_2}}}{3}.
\end{equation}

\subsection{Electrons}

The calculation of the electron optical matrix elements in Eq.~(\ref{A_via_f}) is produced numerically.
The resulting times of radiative intraband transitions are shown in Table~\ref{tab:electron_times}
for QD diameter 3.1~nm.
An empty space in the table means that the corresponding transition is either forbidden by symmetry or is too slow.
\begin{table}[!h]
\caption{\label{tab:electron_times}
        The times of intraband optical transitions of electrons.
        QD diameter is 3.1~nm}
        \begin{tabular}{c|c|c|c|c|c|c|c|c}
            From $\setminus$ To & $1e_0$ & $2e_0$ &  $3e_0$ & $1e_1$ & $2e_1$ & $4e_0$ & $3e_1$ & $1e_2$
            \\
            \hline
            $2e_0$     & 2.4~$\mu$s & --- & --- & --- & --- & --- & --- & ---
            \\
            \hline
            $3e_0$     &  & 0.71~$\mu$s & --- & --- & --- & --- & --- & ---
            \\
            \hline
            $1e_1$     & 0.04~$\mu$s &  &  & --- & --- & --- & --- & ---
            \\
            \hline
            $2e_1$     &  & 0.03~$\mu$s & 0.32~$\mu$s &  & --- & --- & --- & ---
            \\
            \hline
            $4e_0$     & 19.6~$\mu$s &  &  &  &  & --- & --- & ---
            \\
            \hline
            $3e_1$     & 0.8~$\mu$s & 0.03~$\mu$s &  &  &  &  & --- & ---
            \\
            \hline
            $1e_2$     &  & 4.4~$\mu$s &  &  &  &  &  & ---
            \\
            \hline
            $5e_0$     &  &  &  &  &  & 0.17~$\mu$s &  &
            \\
            \hline
        \end{tabular}
\end{table}

\subsection{Holes}

Due to the combination of spherical QD shape and spherical approximation to the hole states,
there are strong selection rules for optical intraband transitions of holes,
which are listed in Table~\ref{tab:rules}.
\begin{table}[!h]
\caption{\label{tab:rules}
        Selection rules for optical intraband transitions}
        \begin{tabular}{c|c|c|c}
             & $hl_0$ & $hh_{F_2}$ & $hm_{F_2}$
            \\
            \hline
            $hl_0$     & $\times$ & $\times$ & $F_2=1$
            \\
            \hline
            $hh_{F_1}$  & $\times$ & $F_1=F_2+1$ & $F_1=F_2$
            \\
            \hline
            $hm_{F_1}$  & $F_1=1$ & $F_1=F_2$   & $F_1=F_2+1$
            \\
            \hline
        \end{tabular}
\end{table}

The times of radiative intraband transitions of holes are given in Table~\ref{tab:hole_times}.
\begin{table}[!h]
\caption{\label{tab:hole_times}
        The times of intraband optical transitions of holes.
        QD diameter is 3.1~nm.
        Sign ``$\times$'' corresponds to the transitions forbidden by the selection rules}
        \begin{tabular}{c|c|c|c|c|c|c|c|c}
            From $\setminus$ To & $1hm_1$ & $1hh_1$ & $1hm_2$ & $1hh_2$ & $1hm_3$ & $2hm_1$ & $1hh_3$ & $1hm_4$
            \\
            \hline
            $1hh_1$     & 1.2~$\mu$s & --- & --- & --- & --- & --- & --- & ---
            \\
            \hline
            $1hm_2$     & 1.3~$\mu$s & $\times$ & --- & --- & --- & --- & --- & ---
            \\
            \hline
            $1hh_2$     & $\times$ & 1.1~$\mu$s & 20~$\mu$s & --- & --- & --- & --- & ---
            \\
            \hline
            $1hm_3$     & $\times$ & $\times$ & 0.55~$\mu$s & $\times$ & --- & --- & --- & ---
            \\
            \hline
            $2hm_1$     & $\times$ & 0.12~$\mu$s & 0.41~$\mu$s & $\times$ & $\times$ & --- & --- & ---
            \\
            \hline
            $1hh_3$     & $\times$ & $\times$ & $\times$ & 0.38~$\mu$s & 9.9~$\mu$s & $\times$ & --- & ---
            \\
            \hline
            $1hm_4$     & $\times$ & $\times$ & $\times$ & $\times$ & 0.33~$\mu$s & $\times$ & $\times$ & ---
            \\
            \hline
            $1hl_0$     & 6.0~ns & $\times$ & $\times$ & $\times$ & $\times$ & 0.56~$\mu$s & $\times$ & $\times$
            \\
            \hline
        \end{tabular}
\end{table}

\section{Confined hole energy relaxation due to multiphonon transitions}
\label{sec:multiphonon}

Although most of the gaps between the calculated energy levels in the valence band
appear to be large enough ($>100$~meV) to suppress intraband relaxation processes assisted
by emission of a single phonon in QDs with diameter in the range of $2 \div 4$~nm.
However, nonradiative transitions in which the valence-band hole emits several
optical phonons (energy of optical phonon in Si is about 60 meV) are possible.
The rate of such multiphonon intraband energy relaxation
is governed by the Huang-Rhys factor $S$~\cite{HuangRhys, Abakumov_book, Goupalov_PRB_2005}.
In case of $S \ll 1$ the rate of transition assisted by emission of $p$ phonons, $W_p$,
is proportional to the factor of the order of $S^p/p!$ compared to the single-phonon transition rate.
Below we present a calculation of the Huang-Rhys factor
for intraband transitions of valence-band holes confined in Si/SiO$_2$ nanocrystals.

In non-polar Si interaction of phonons with holes is determined by the deformation potential.
Interaction of electrons with optical phonons via the deformation potential is forbidden by the crystal symmetry.
Since optical phonons in Si are almost dispersionless, the system under study can be
described by multi-mode Huang-Rhys model, similar to the one-mode model~\cite{Abakumov_book, Goupalov_PRB_2005}.

We consider two hole levels of size quantization $|i\rangle$ and $|j\rangle$ in the QD potential
with energies $\varepsilon_i$, $\varepsilon_j$  ($\varepsilon_j>\varepsilon_i$) interacting with
optical phonons characterized by frequency $\omega_0$, wavevector $\bm q$ and polarisation $\sigma$.
The Hamiltonian of the hole interaction with optical phonons in Si
is given by~\cite{BirPikus_book}:
\begin{equation}
\label{optham}
    H_{\rm e-ph}({\bf u})
    = \frac{2}{\sqrt{3}} \frac{d_0}{a_0}
    \left(
        u_x \, \{J_y J_z\} +u_y \{J_z J_x\} +u_z \{J_x J_y\}
    \right),
\end{equation}
where $2\{J_\alpha J_\beta \}=J_\alpha J_\beta + J_\beta J_\alpha $,
$a_0$ is the lattice constant,
$d_0$ is the interaction constant.
The value of $d_0$ in silicon confirmed by experimental data is 27~eV~\cite{Cardona_PRB_1987, Cardona_PSSB_1984}.
But when spin-orbit interaction is neglected (that is the approximation we use)
and holes are characterized by momentum $J=1$,
one can easily find that the value of $d_0$ should be multiplied by 3 (compared to the case of $J=3/2$).
So $d_0=81$~eV is used in our consideration.

The continuous field of the relative atomic displacement for a given bulk phonon mode is
\begin{equation}
    {\bf u}_{{\bf q},\sigma}({\bf r})
    = \sqrt{\frac{\hbar}{2 \rho_0 \omega_0 V}}
    \left(
          b^{\dag}_{\bf q,\sigma} \bm e_{\bf q,\sigma}^* e^{-i{\bf qr}}
        + b_{\bf q,\sigma} \bm e_{\bf q,\sigma} e^{i {\bf qr}}
    \right),
\end{equation}
where $\rho_0$ is the reduced mass density,
and the polarization vectors $\bm e_{\bf q,\sigma}$ can be chosen to be purely real.
It is convinient to introduce following  matrix elements:
\begin{gather}
 M_{i,j}(\bm q,\sigma)=\sqrt{\frac{\hbar}{2 \rho_0\omega_0 V}} \langle i | H_{\rm e-ph}(\bm e_{\bm q,\sigma} ) \, e^{i {\bf qr}}|j \rangle \:.
\end{gather}

Electron-phonon interaction induces polaron shifts to energy levels under consideration.
Taking these shifts into account, the number of phonons emitted is given by
\[
 p=\frac{\varepsilon_j-\varepsilon_i}{\omega_0}+\sum\limits_{{\bf q}, \sigma}\frac {|M_{i,i}({\bf q}, \sigma)|^2
 -|M_{j,j}({\bf q}, \sigma)|^2}{\omega_0^2}\:.
 \]
The Huang-Rhys factor is given by
\begin{equation}
S = \sum\limits_{{\bf q}, \sigma} \frac1{\omega_0^2}\left |M_{i,i}(\bm q,\sigma)-M_{j,j}(\bm q,\sigma)\right|^2\:.
\label{huangrhys}
\end{equation}

\begin{figure}
    \includegraphics[width=12cm]{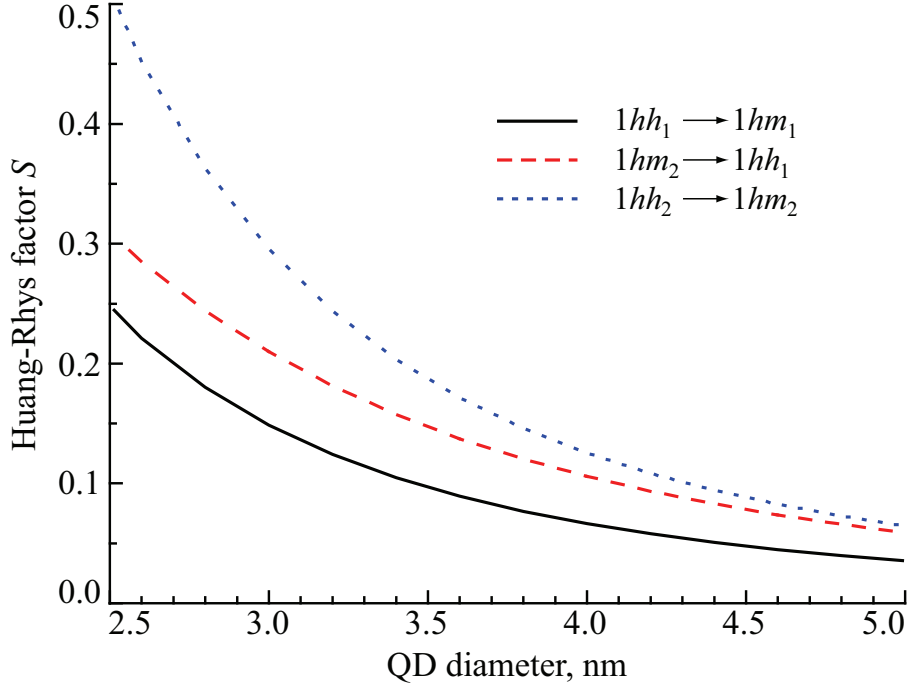}
\caption{
    \label{fig:S_HR}
    Huang-Rhys factor $S$ vs. quantum dot diameter for multiphonon hole transitions between the first four levels
}
\end{figure}

The calculated values of $S$ for the transitions between the four lowest hole levels are presented in Fig.~\ref{fig:S_HR}.

The phonon factor corresponding to the number of phonons that should be emitted for transitions between neighboring levels ($p \ge 2$)
is of the order of $10^{-2}$~\cite{Prokofiev_JoL_2006}.
Keeping in mind that the characteristic time of the one-phonon transition is 1~ps,
we conclude that the rate of multiphonon energy relaxation of hot holes is of the order of $10^{10}$~s$^{-1}$.

\section{Conclusions}
\label{sec:conclusions}

For spherical silicon nanocrystals in SiO$_2$ under consideration
the estimated rate of multi-phonon hole transitions ($10^{10}$~s$^{-1}$)
is higher than both electron and hole optical intraband transitions rates,
lying in the range of $10^5 \div 10^8$~s$^{-1}$.
All the intraband relaxation mechanisms considered above are much faster
than optical recombination of confined carriers
(whose rate is not higher than $10^4$~s$^{-1}$~\cite{Moskalenko_PRB_2007}).
Thus, when considering radiative recombination of confined excitons,
one may assume most of the carriers to be in the ground states of QD.

This work was supported in part by DoD under contract No.~W912HZ-06-C-0057.


\begin{thebibliography}{99}
\bibitem{Kenyon_PQE_2002}
A.~J.~Kenyon,
Prog. Quantum Electron. \textbf{26}, 225 (2002).

\bibitem{Pavesi_Nature_2000}
L.~Pavesi, L.~Dal Negro, C.~Mazzoleni, G.~Franzo, F.~Priolo,
Nature \textbf{408}, 440 (2000).

\bibitem{Timmerman_Nature_2008}
D.~Timmerman, I.~Izeddin, P.~Stallinga, I.~N.~Yassievich, T.~Gregorkiewicz,
Nature Photonics \textbf{2}, 105 (2008).

\bibitem{Moskalenko_PRB_2007}
A.~S.~Moskalenko, J.~Berakdar,  A.~A.~Prokofiev, I.~N.~Yassievich,
Phys. Rev.~B \textbf{76}, 085427 (2007).

\bibitem{BirPikus_book}
G.~L. Bir and G.~E. Pikus,
\textit{Symmetry and Strain-Induced Effects in Semiconductors}
(Nauka, Moscow 1972).

\bibitem{Allan_PRB_2002}
G.~Allan, C.~Delerue,
Phys. Rev. B \textbf{66}, 233303 (2002).

\bibitem{Goupalov_PRB_2003}
S.~V.~Goupalov,
Phys. Rev. B \textbf{68}, 125311 (2003).

\bibitem{Thraendhardt_PRB_2002}
A.~Thr\"{a}ndhardt, C.~Ell, G.~Khitrova, and H.~M.~Gibbs,
Phys. Rev. B {\bf 65}, 035327 (2002).

\bibitem{HuangRhys}
K.~Huang and A.~Rhys,
Proc. Roy. Soc. London A {\bf 204}, 406 (1950);
K.~Huang,
Scientia Sinica {\bf 24}, 27 (1981).

\bibitem{Abakumov_book}
V.~N. Abakumov, V.~I. Perel, and I.~N. Yassievich,
in \textit{Nonradiative Recombination in Semiconductors},
edited by V.~M. Agranovich and A.~A. Maradudin,
Modern Problems in Condensed Matter Sciences
Vol.~33, Elsevier, Amsterdam, 1991.

\bibitem{Goupalov_PRB_2005}
S.~V. Goupalov,
Phys. Rev. B {\bf 72}, 073301 (2005).

\bibitem{Cardona_PRB_1987}
L.~Brey, N.~E. Christensen, and M.~Cardona,
Phys. Rev. B {\bf 36}, 2638 (1987).

\bibitem{Cardona_PSSB_1984}
A.~Blacha, H.~Presting, and M.~Cardona,
Phys. Status Solidi~B {\bf 126}, 11 (1984).

\bibitem{Prokofiev_JoL_2006}
A.~A.~Prokofiev, A.~S.~Moskalenko, I.~N.~Yassievich,
J.~of Lumin. \textbf{121}, 222 (2006).
\end{thebibliography}
\end{document}